\documentclass[prd,amsmath,amssymb,aps,superscriptaddress,twocolumn,preprintnumbers,nofootinbib]{revtex4-2}

\usepackage{epsfig}
\usepackage{url}
\usepackage{hyperref}

\usepackage[normalem]{ulem} 

\usepackage{latexsym}
\usepackage{epsfig}
\usepackage{amsmath}
\usepackage{amssymb}
\usepackage{wasysym}
\usepackage{graphicx}
\usepackage{dcolumn}
\usepackage{verbatim}
\usepackage{enumerate,mdwlist}
\usepackage{soul}

\usepackage[titletoc]{appendix}

\usepackage{amsfonts}
\usepackage{fancyvrb} 
\usepackage{tikz} 
\usetikzlibrary{calc}
\usepackage[export]{adjustbox}

\usepackage[normalem]{ulem}

\usepackage[inline, final]{showlabels}
\usepackage{tensor}

\usepackage{listings}
\usepackage[dvipsnames]{xcolor}
\usepackage{fontawesome5}

\definecolor{comment_red}{rgb}{0.5, 0, 0}
\newcommand{\ii}{\text{i}}
\newcommand{\di}{\text{d}}

\makeatletter
\g@addto@macro\bfseries{\boldmath}  
\makeatother

\renewcommand{\v}[1]{\boldsymbol{#1}}

\renewcommand{\d}[2]{\frac{\text{d} #1}{\text{d} #2}}

\newcommand{\ba}{\begin{eqnarray}}
\newcommand{\ea}{\end{eqnarray}}

\newcommand{\glow}{\texttt{GLoW}}
\newcommand{\dingo}{\textsc{DINGO}}
\newcommand{\uldingo}{unlensed-DINGO}
\newcommand{\lensdingo}{lensed-DINGO}

\definecolor{grey}{rgb}{0.4,0.4,0.4}
\definecolor{dullmagenta}{rgb}{0.4,0,0.4}
\definecolor{darkblue}{rgb}{0,0,0.4}
\definecolor{midblue}{rgb}{0,0,0.5}
\definecolor{midred}{rgb}{0.5,0,0}
\definecolor{orange}{rgb}{1,0.5,0}
\definecolor{lightbrown}{rgb}{0.75,0.5,0.25}
\definecolor{tan}{cmyk}{0.14,0.42,0.56,0}
\definecolor{djunglegreen}{cmyk}{0.99,0,0.52,0}
\definecolor{lightgreen}{rgb}{0,1,0}
\definecolor{olivegreen}{cmyk}{0.64,0,0.95,0.40}
\definecolor{midgreen}{rgb}{0.0,0.675,0.0}
\definecolor{darkgreen}{rgb}{0,0.5,0}
\definecolor{ceruleanblue}{rgb}{0.0, 0.2, 0.7}
\definecolor{burgundy}{rgb}{0.5, 0.0, 0.13}
\definecolor{hvred}{RGB}{186,12,47}

\hypersetup{
    colorlinks=true,
    linkcolor=ceruleanblue,
    filecolor=ceruleanblue,      
    urlcolor=midblue,
    citecolor=burgundy,
}

\usepackage[all]{xy} 
\usepackage{amsfonts}

\makeatletter
\def\l@subsubsection#1#2{}
\makeatother

\begin{document} 


\title{Accelerated inference of microlensed gravitational waves with machine learning}

\author{Marienza Caldarola}
\email{marienza.caldarola@csic.es}
\affiliation{Instituto de F\'isica Te\'orica UAM-CSIC, Universidad Aut\'onoma de Madrid, Cantoblanco, 28049 Madrid, Spain}

\author{Srashti Goyal}
\email{srashti.goyal@aei.mpg.de}
\affiliation{Max Planck Institute for Gravitational Physics (Albert Einstein Institute) \\
Am Mühlenberg 1, D-14476 Potsdam-Golm, Germany}

\author{Nihar Gupte}
\email{nihar.gupte@aei.mpg.de}
\affiliation{Max Planck Institute for Gravitational Physics (Albert Einstein Institute) \\
Am Mühlenberg 1, D-14476 Potsdam-Golm, Germany}

\author{Stephen R. Green}
\email{stephen.green2@nottingham.ac.uk}
\affiliation{School of Mathematical Sciences, University of Nottingham, Nottingham NG7 2RD, United Kingdom}

\author{Miguel Zumalac\'arregui}
\email{miguel.zumalacarregui@aei.mpg.de}
\affiliation{Max Planck Institute for Gravitational Physics (Albert Einstein Institute) \\
Am Mühlenberg 1, D-14476 Potsdam-Golm, Germany}

\begin{abstract}
Gravitational waves (GWs) within the LIGO-Virgo-KAGRA sensitivity band can be microlensed by stellar and intermediate-mass black holes, producing a frequency-dependent modulation of the signal amplitude. Microlensing analyses, however, are costly due to the increased dimensionality of the parameter space and waveform computation time.
As a proof of concept, we show that the deep-learning-based framework Deep Inference for Gravitational-Wave Observations (DINGO), which employs a simulation-based inference approach to estimate posterior distributions, can perform efficient parameter inference for GW microlensing by an isolated point-mass lens. Using simulated microlensed GW signals, we train a \lensdingo{} network and compare its performance with traditional Bayesian parameter estimation carried out with Bilby. Our framework can be used to rapidly identify microlensed events in large GW catalogs. When the \lensdingo{} network is combined with importance sampling, we find that although sample efficiencies are somewhat reduced compared to the \uldingo{} network, owing to the richer structure of microlensed signals, it still achieves $\mathcal{O}(10\times)$ speed-up relative to Bilby. We further show that this framework is useful to efficiently estimate the background Bayes-factor distribution, which is crucial for assessing the significance of candidate lensed events. However, for foreground (lensed) events, the sampling efficiency can sometimes drop when analysed with the \uldingo{} network, providing a diagnostic indicator of out-of-distribution data. Our approach can be straightforwardly generalised to more complex and realistic lens models, enabling detailed studies of microlensed GWs.
\end{abstract}

\date{\today}

\maketitle

\section{Introduction}\label{sec:intro}

Gravitational waves (GWs) are ripples in spacetime generated by the acceleration of massive compact objects, such as merging black holes or neutron stars. Since their first detection by LIGO and Virgo Collaboration in 2015~\cite{Abbott:2016blz}, they have become a powerful probe of gravity, astrophysics, and cosmology~\cite{Abbott2016_TGR_GW150914,Isi:2019aib,Cardoso:2017cfl,LIGO2025_TGR_GWTC3,Abbott2023_GWTC3_Population,Abbott2017_HubbleConstant,Abbott2023_CosmicExpansion_GWTC3}.

Like any other signal, GWs travelling through the Universe are deflected and distorted by the gravitational fields along their path~\cite{Schneider1999-tk}. Gravitational lensing offers a powerful probe of cosmology, astrophysics, and fundamental physics, with applications such as constraining the distribution of dark matter~\cite{Clowe:2006eq,Planck:2018lbu,Vegetti:2023mgp}.
Weak lensing produces small distortions that trace the large-scale matter distribution, while strong lensing constrains the mass profiles of galaxies and clusters~\cite{Bartelmann:1999yn,Treu:2010uj,Natarajan:2024iqm}. Lensing phenomena can enable precision tests of general relativity~\cite{Reyes:2010tr,Collett:2018gpf,Goyal:2023uvm, Goyal:2020bkm} and cosmological parameters~\cite{H0LiCOW:2019pvv,Treu:2022aqp,Jana:2022shb}.
Gravitational lensing phenomena are essential to correctly interpret observations across the electromagnetic spectrum~\cite{Smith:2025axx}. For GWs, in addition to being magnified and occasionally split into multiple images, lensing can produce novel effects due to the signal's finite wavelength, such as diffraction and interference~\cite{Takahashi:2003ix,Leung:2023lmq}. Accurately modeling these effects is crucial: if not accounted for, lensing can bias the posteriors and introduce systematic errors in source and cosmological inference.
Most importantly, diffraction may be the smoking gun for small-scale objects and distant magnified GWs~\cite{Diego:2019lcd,Cheung:2020okf,Yeung:2021chy,Mishra:2021xzz,Oguri:2022zpn,Mishra:2023ddt,Shan:2024min,Lo:2024wqm,Seo:2025dto,Smith:2025axx,Su:2025xry}.

GW lensing can be described in two regimes: the geometric optics (GO) regime, where the GW wavelength is much smaller than the characteristic gravitational radius of the lens, and the wave-optics (WO) regime, where the GW wavelength is comparable or bigger than the characteristic size of the lens, where diffraction and interference become significant. We focus on the WO regime, as its characteristic interference patterns imprint distinctive, frequency-dependent modulations on the waveform, as clear signatures of lensing by compact objects. These effects are often combined, with a macroscopic lens (a galaxy or cluster) magnifying the GW, and small-scale objects (stars, remnants) diffracting it; these cases are referred to as macrolensing and microlensing, respectively~\cite{Mishra:2023ddt}.

Lensing computations are challenging in the WO regime, as they require solving highly oscillatory integrals~\cite{Feldbrugge:2019fjs,Tambalo:2022plm}. Recent advances in algorithms~\cite{Shan:2022xfx,Yeung:2024pir,Villarrubia-Rojo:2024xcj} and parameter estimation (PE)~\cite{Wright_2022,Cheung:2024ugg} now enable the search for these signatures among GW events. Despite this progress, inference becomes more costly due to the computational cost of diffraction integrals, and the extended parameter space. Intriguing candidates for lensed GWs have been reported~\cite{GW231123_independent,LIGOScientific:2025rsn,Janquart:2023mvf}, although none yet represents an unambiguous detection~\cite{LIGOScientific:2021izm,LIGOScientific:2023bwz,LIGOScientific:2025cwb}. Nevertheless, lensed GWs are inevitable due to the rapidly increasing number of observed events, warranting a search among the $\mathcal{O}(100/{\rm yr})$ in the current run, the fourth observing run (O4), which will reach $\mathcal{O}(10^5/{\rm yr})$ with next-generation detectors~\cite{Broekgaarden:2023rta}. Many forecasts agree that lensed GWs will be commonplace in the next observing run O5~\cite{Ng:2017yiu,Li:2018prc,Smith:2022vbp,Wierda:2021upe}, but they have to be identified among thousands of signals observed every year.

Given the inevitability of lensed signals within the growing data volume, the challenges of WO computations, and the complexity of traditional analysis, we are motivated to develop a machine learning framework for fast and accurate PE of lensed GW signals. 
Traditional Bayesian inference methods, such as nested sampling~\cite{Skilling:2006gxv, Speagle:2019ivv, Ashton:2018jfp}, are computationally expensive and often prohibitively time consuming for large datasets. Deep learning techniques have emerged as a powerful alternative, enabling rapid and efficient PE in GWs analyses~\cite{Santoliquido:2025lot,Qin:2025mvj}. Recent studies have also explored machine-learning-based inference for GW lensing parameters using various methodologies, pursuing similar objectives to this work. Examples include the use of conditional variational autoencoders~\cite{Bada-Nerin:2024wkn}, normalizing flows for stochastic lensing ~\cite{Su:2025xry}. While preliminary, our approach demonstrates the potential for extending machine-learning-based microlensing inference to real data in future GW analyses. 

In this study, we model lensing effects on GW signals by computing the amplification factor using 
Gravitational Lensing of Waves (\glow{}), a precise and flexible code that calculates
frequency-dependent amplification for general lens configurations. PE is then performed using 
the Deep Inference for Gravitational-wave Observations (DINGO) framework~\cite{Dax:2021tsq,
Green:2020hst, Green:2020dnx, Dax:2022pxd}. DINGO has been reviewed by the LIGO-Virgo-KAGRA (LVK) Collaboration, used for eccentricity estimates~\cite{Gupte:2024jfe}, and extended to work on binary neutron star systems~\cite{Dax:2024mcn}. Recently, it has also been applied to overlapping GW signals~\cite{hu2025hierarchicalsubtractionneuraldensity}.

The paper is organised as follows. In Sec.~\ref{sec:form} we introduce the gravitational lensing of GWs. In Sec.~\ref{sec:Bayes_theory} we address the Bayesian PE methods, while in Sec.~\ref{sec:dingo} we review the DINGO framework and mention our train setup in Sec.~\ref{sec:train}. In Sec.~\ref{sec:results} we present our results and provide conclusions in Sec.~\ref{sec:concl}.

\section{GW lensing}\label{sec:form}
Gravitational lensing in the weak field limit modulates the GW strain by a frequency-dependent amplification factor~\cite{Schneider1999-tk}, 
\begin{equation}
    \label{F_def}
    F(f) \equiv \frac{\tilde{h}_\text{L}(f)}{\tilde{h}_0(f)}  \ ,
\end{equation}
where $\tilde{h}_0(f)$ and $\tilde{h}_\text{L}(f)$ denote the frequency-domain unlensed and lensed strains, respectively. The amplification factor is calculated from the diffraction integral over the lens plane $\v{x}$,
\begin{equation}\label{eq:Fw_def}
    F(w) = \frac{w}{2\pi \ii}\int\di^2 x\exp\Big(\ii w({ \phi}(\v{x}, \v{y})- \phi_0)\Big)\ ,
\end{equation}
where $\phi(\v{x}, y)= \frac{1}{2}|\v{x}-\v{y}|^2 - \psi(\v{x})$  is the Fermat potential, $\phi_0$ is its minimum and $\v{y}$ is the source position, rescaled by an arbitrary length scale $\xi_0$. The lensing potential $\psi(x)$ depends on the projected mass density of the lens (through a Poisson equation) and the dimensionless frequency,
\begin{equation}
w \equiv 8\pi GM_{Lz}f  \approx 1.24 \left(\frac{M_{Lz}}{100 M_\odot}\right)\left(\frac{f}{100 Hz}\right),    
\end{equation}
is a scaled version of GW signal frequency $f$ using the effective lens mass $M_{Lz} \equiv \frac{\xi_0^2 (1+z_L) D_{\rm S}}{4G D_{\rm L}D_{\rm LS}}$, 
where $D_{\rm S}$ is the angular distances to the source, $D_{\rm L}$ to the lens and $D_{\rm LS}$ between lens and source. We have set $c = 1$.

In the GO limit ($w\gg1$), the amplification factor gets a contribution only from the stationary points of the Fermat potential, where the images are formed. However, in the WO regime ($w\sim1$), the full diffraction integral needs to be evaluated, which for most types of lenses cannot be done analytically. A set of numerical algorithms is packaged in \glow{} to evaluate this integral efficiently for various lens models~\cite{Villarrubia-Rojo:2024xcj}.
\footnote{The \glow{} code is publicly available at \url{https://github.com/miguelzuma/GLoW_public}
} The diffracted signal thus encodes the information about the lens properties and allows one to probe dark and baryonic matter in the Universe~\cite{Urrutia:2021qak,Tambalo:2022wlm,Savastano:2023spl,GilChoi:2023qrz,Zumalacarregui:2024ocb,Vujeva:2025nwg,Brando:2024inp}.

Many models are used to describe the mass distribution of lenses~\cite{Keeton:2001ss}. In the simplest case, point lens models, representing compact objects such as planets, stars, or black holes, allow for analytical treatment of the lensing properties~\cite{Jow:2020rcy}. More complex astrophysical lenses, such as galaxies or clusters, are described using extended profiles like the Singular Isothermal Sphere~\cite{hinshaw1987gravitational} and Navarro-Frenk-White~\cite{Navarro:1996gj}. Even then, realistic representations require the superposition of many such simple lenses, often including additional deformations (ellipticity, multipolar structure). The $F(w)$ computations become much more expensive for extended lenses, and machine learning can be particularly helpful in such cases.

As a proof of principle, we will consider only the isolated point-mass lens, neglecting an environmental potential and additional objects. The point lens lensing potential takes the form $\psi(\v{x}) = \ln x$. Conveniently, choosing the scale  $\xi_0 = R_E$, where $R_E \equiv \sqrt{\frac{4GM_L D_{\rm L} D_{\rm LS}}{D_{\rm S}}} \approx 0.14{\rm pc} \left(\frac{M_L}{100M_\odot}\right)^{1/2}\left(\frac{D_{\rm L}D_{\rm LS}D_{\rm S}^{-1}}{1{\rm Gpc}}\right)^{1/2}$ as the Einstein radius of the point-mass lens, yields $M_{Lz} = M_L(1+z_L)$, where $M_L$ is the mass of the lens and $z_L$ is its redshift. $M_{Lz}$ is equivalent to the ``redshifted lens mass''. Effectively, the diffraction by a point-mass lens on the GW waveform is given by a frequency-dependent amplification factor $F(f)$ that depends on two parameters: $M_{Lz}$ and $y$, where $y$ is the impact parameter of the source in units of the Einstein radius.
It can be computed analytically as a hypergeometric series expansion, for which several approximations speed up the computation in different regions of the parameter space (Sec.~V A in \cite{Villarrubia-Rojo:2024xcj}).

In the GO limit, which holds for $w\gg1$, the lensed waveform is approximately equivalent to an interference of two images. The amplification factor can be written in GO approximation as
\begin{equation}
F(f) \approx \sqrt{\mu_+}\ + \sqrt {\mu_-}\ \exp(2\pi if\Delta t -\pi/2)\,.
\end{equation}
This gives rise to two images with magnifications $\mu_\pm = \frac{1}{2} \left| \frac{y^2 + 2}{y \sqrt{y^2 + 4}} \pm 1 \right|$ (defined positive), and separated by a time delay,
\begin{align}
\Delta t &= 
4 G M_{Lz}
\left[
\frac{y}{2}\sqrt{y^2 + 4}
+ \ln\!\left(\frac{\sqrt{y^2 + 4} + y}{\sqrt{y^2 + 4} - y}\right)
\right]
\; \nonumber \\
&\approx\;
4 \mathrm{ ms}
\left(\frac{M_{Lz}}{1000\,M_\odot}\right),
\quad \mathrm{at } \quad y = 1.
\end{align}

\begin{figure}[!t]
\centering
\includegraphics[width = 0.48\textwidth]{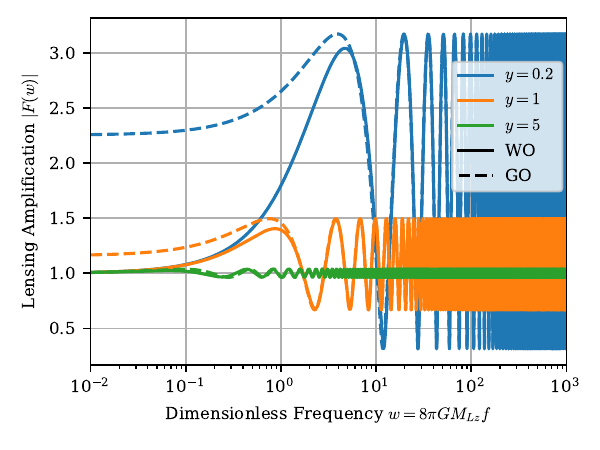}
\caption{Lensing \label{fig:amplif_factor} amplification factor $F(w)$ for a point-mass lens at different impact parameters $y$, in units of the Einstein radius $R_E$. The WO regime is near $w\sim1$, whereas for $w\gg1$ the GO approximation is valid. The smaller impact parameters modulate the waveform much more than the larger impact parameters, such that for  $y\gg1$ the lensing effect is negligible $F(w) \sim 1$. }
\end{figure}

An example of $F(w)$ for a point lens, along with the GO approximation, using different impact parameters $y$ is given in Fig.~\ref{fig:amplif_factor}. Notice that for high frequencies $w\gg1$ and/or lens masses, the $F(w)$ is very oscillatory; this is where GO is a good approximation. On the other hand, for $w \sim 1$ the GO approximation fails and one needs to consider the full WO amplification factor. As expected, the smaller impact parameters have a more pronounced effect on the waveform than the larger impact parameters. For example, for $y\sim5$, the lensing amplification is negligible with $F(w) \sim 1$, irrespective of the mass of the lens. The LIGO-Virgo detectors are most sensitive in the frequency range $f \in [10,1000]$ Hz, which corresponds to the WO lensing regime for isolated point masses with $M_{Lz} \in [10,10^4]M_\odot$.

\section{Bayesian PE}
\label{sec:Bayes_theory}    
In the context of GW analysis, the observed detector data are given by $d(t)=h(t,\theta)+n(t)$, where $h(t,\theta)$ is a GW strain template for the waveform given the source parameters $\theta$ and $n(t)$ the detector noise. The Bayes theorem is used to infer the properties of an observed GW source, 
\begin{equation}
    \label{Bayes_theorem}
    p(\theta|d,H) = \frac{p(d|\theta,H) p(\theta|H)}{p(d|H)}\ ,
\end{equation}
where $p(\theta|d,H)$ is the posterior probability distribution of the parameters $\theta$ given the data $d$ and the hypothesis $H$,  $p(\theta|H)$ is the prior, and $p(d|\theta, H)$ is the likelihood. The denominator $p(d|H)$ is Bayesian evidence $Z_H$, which is useful for model selection. Given two hypotheses, in our case lensed $H_L$ and unlensed $H_U$, the Bayes factor is the ratio of evidences $B^L_U= Z_L/Z_U=P(d|H_L)/ P(d|H_U)$. This ratio quantifies the preference of one hypothesis over the other, considering only the data. 
 
In GW astronomy,  a Gaussian noise likelihood function is generally assumed, which in the frequency domain takes the form~\cite{Thrane_2019}
\begin{equation}
    p(d|\theta) \propto \exp{\Big(-2\int_0^\infty\,df\,\frac{|\tilde d(f)-\tilde h(f,\theta)|^2}{S_n(f)} \Big)}\,,
\end{equation}
where $\tilde h(f,\theta)$ represents the GW signal in the frequency domain for a given set of parameters $\theta$, projected onto the detector antenna response pattern, while $S_n$ is the power spectral density (PSD) of the detector noise. For any signal $\tilde h(f)$, its optimal signal-to-noise ratio (SNR) is given by $\rho_{opt} =\left[ 4 \int_{0}^{\infty} \frac{|\tilde{h}(f)|^2}{S_n(f)} \, df \right]^{1/2}$.

To estimate the posterior, one typically relies on evaluating the likelihood explicitly for many parameter samples from the prior while using efficient sampling techniques like nested sampling~\cite{Skilling:2006gxv}. In GW astronomy, often one uses the Bilby~\cite{Ashton:2018jfp} code, which is a GW specific interface to many 
nested sampling implementations. The most popular of these is \texttt{dynesty}~\cite{Speagle:2019ivv}.

Given the 15-dimensional parameter space for a binary black hole, and an additional two parameters in the case of the isolated point lens, sampling turns out to be computationally expensive taking $\mathcal{O}$(days). It is especially challenging for constructing a background distribution of Bayes factors, which requires a large number of unlensed injections (e.g., currently $\sim$$400$ injections for $3\sigma$ deep background) to estimate the significance of a lensed event candidate. 
These methods are not feasible given the expected number of GW events, which will rise as detector sensitivities improve, reaching $\mathcal{O}(10^5)$ binary black hole events per year for next-generation detectors~\cite{EventHorizonTelescope:2019dse}. This motivated us to use the simulation-based inference tool DINGO~\cite{Green:2020hst,Dax:2021tsq}.

\section{DINGO framework}\label{sec:dingo}
DINGO\footnote{\url{https://github.com/dingo-gw/dingo}} utilizes neural posterior estimation (NPE) with normalizing flows to model the posterior distribution $p(\theta|d)$. The flow $f_\phi$ learns an invertible transformation from a
simple base distribution $p(\mathbf{u})$ to the complex posterior, conditioned on the
data: $\theta = f_{\phi}(\mathbf{u}; d)$ \cite{papamakarios2021normalizingflowsprobabilisticmodeling}. The resulting probability density,
$q_{\phi}(\theta|d)$, is given by the change of variables formula,
\begin{equation}
q_{\phi}(\theta|d) = p\left(\mathbf{u}=f_{\phi}^{-1}(\theta; d)\right)
\left| \det J_{f_{\phi}^{-1}}(\theta; d) \right|\,,
\end{equation}
where $J_{f_{\phi}^{-1}}$
is the Jacobian of the inverse transformation. The network is trained by
minimizing the forward Kullback-Leibler divergence between its output distribution
$q_{\phi}(\theta|d)$ and the true posterior $p(\theta|d)$. This is achieved by
minimizing the negative log-likelihood on a training set of simulated
signal-parameter pairs $(\theta_i, d_i)$, generated by injecting GW signals into
stationary Gaussian noise realizations. The loss function is the negative log-likelihood, averaged over the
joint distribution of parameters and data, $p(\theta, d)$,
\begin{equation}\mathcal{L}(\phi) =
 \mathbb{E}_{p(\theta, d)} [-\log q_{\phi}(\theta|d)]\,.
\end{equation}
Since the true
posterior is unknown, this loss is expressed as an expectation over the prior
$p(\theta)$ and the likelihood $p(d|\theta)$, which can be approximated using a
Monte Carlo average over training samples,
\begin{equation}
\mathcal{L}(\phi) \approx - \frac{1}{N} \sum_{\substack{\theta^{(i)} \sim p(\theta)\\d^{(i)} \sim p(d|\theta^{(i)})}}\log
q_{\phi}(\theta^{(i)}|d^{(i)})\,,
\end{equation}
where $N$ is the number of samples in the mini-batch.
This training process allows the network to
perform amortized inference, generating posterior samples for new events in
seconds, without evaluating any likelihoods explicitly. This approach is 
called ``simulation-based inference''.

To simplify the learning task for the
primary NPE network, we infer the coalescence time in each detector using the
group-equivariant neural posterior estimation (GNPE) algorithm
\cite{Dax:2021myb,Dax:2021tsq}. This approach standardizes the time of arrival
of the signal across the detector network. Although this requires training a
separate neural network, it simplifies the data by aligning coalescence times,
allowing the main NPE network to focus more effectively on the remaining
physical parameters of the source.

DINGO also allows for importance sampling (IS), in which the output samples are corrected by explicit likelihood evaluation~\cite{Dax:2022pxd}. Given our target distribution $p(\theta|d) \propto p(d|\theta)p(\theta)$ and a set of $n$ samples from the proposal distribution,
each sample $\theta_i \sim q(\theta|d)$ is assigned an importance weight,
\begin{equation}
    w_i=\frac{p(d|\theta_i) p(\theta_i)}{q(\theta_i|d)}\,.
\end{equation}
Since the proposal is unlikely to be perfect, the effective samples size
\begin{equation}
    n_\text{eff} = \frac{(\sum_i w_i)^2}{\sum_i w_i^2} < n\,,
\end{equation}
resulting in a sample efficiency
\begin{equation}
    \epsilon = \frac{n_\text{eff}}{n}\,.
\end{equation}
The sample efficiency provides a measure of proposal quality~\cite{Dax:2022pxd}.  IS requires evaluating $p(d|\theta)p(\theta)$ for all $n$ samples. For sufficiently large $\epsilon$, DINGO-IS requires many fewer likelihood evaluations than conventional sampling techniques, and moreover they can be executed in parallel. The evidence $Z=p(d) = \int d\theta P(d|\theta)P(\theta)$ is estimated from the normalization of the weights,
\begin{equation}
    Z =p(d)=\frac{1}{n}\sum_i w_i\,.
\end{equation}
IS corrects for any mismatch between the proposal distribution and the true posterior and thereby improves the final inference results. The statistical uncertainty of the IS evidence estimator
is controlled by the variance of the importance weights ~\cite{Dax:2022pxd},
\begin{equation}
\mathrm{\sigma^2}(Z) = \frac{1}{n}\,\mathrm{\sigma^2}(w)
= \frac{1}{n}\left(\langle w^2\rangle - \langle w\rangle^2\right).
\end{equation}
Consequently, the relative error on the evidence scales as
\begin{equation}
\frac{\sigma({Z})}{Z} \propto \sqrt{\frac{1}{n_{\mathrm{eff}}}} .
\end{equation}
Hence, for an accurate evidence estimate, the effective sample size should be large. Meaning for lower sampling efficiency, one would need a larger number of proposal samples ($n$) to achieve the desired $n_{\rm eff}$. In all cases considered in this study, a conservative requirement on the effective sample size is imposed ($n_{\rm eff} \gtrsim 500$), ensuring that the resulting evidence estimates and Bayes factors are not dominated by a small number of high-weight samples.

\section{Training Process}\label{sec:train}
For training \lensdingo{} network, we implement a WO lensing transform that multiplies $F(f)$ (see Sec. \ref{sec:form}) to simulated signals on the fly during training. This maximises coverage of the lensing parameters and mitigates overfitting.  As the point lens waveform computation is very fast $\mathcal{O}$(0.1 ms), this is a reasonable approach; however, for more complicated lensed models, lensed waveforms should be computed in advance of training.\footnote{All modifications to include lensing in DINGO have been developed in a branch, available publicly at \url{https://github.com/srashtig/dingo/tree/lensing}} We evaluate the amplification factor on the signals using \glow{} for a point lens in the WO regime (see Sec.~\ref{sec:form}). 

In addition to lensed signals, we also train DINGO without any lensing transform, representing the standard analysis (15 parameters), to compare the results. 
From now on, we will distinguish the two networks as \uldingo{} and
\lensdingo{}. For both networks, we used $10^7$  waveforms in the
frequency domain based on the \texttt{IMRPhenomXPHM} approximant~\cite{Pratten:2020ceb}, sampled at $2048$
Hz, in the range $[20,1024]$ Hz with a duration of 8 sec. The GW signals are injected into the Gaussian
noise, sampled from a fixed PSD of the Hanford, Livingston and Virgo detectors using the data from the third
observing run (O3)~\cite{O3sensitivity}.  We estimate PSDs using the Welch method with 
1024 sec of data and a sampling frequency of 4096 Hz. The training
dataset is generated using the standard prior distributions for usual 15
parameters of the source, see Table~\ref{tab:priors}. The lens parameters are generated for the redshifted
lens mass, $\log_{10}(M_{Lz})\in \rm Uniform [1.0,4.0]$, and the impact parameter, $y\in \rm Uniform [0.1,5.0]$, as shown in Table~\ref{tab:priors}.
\renewcommand*{\thefootnote}{\alph{footnote}}
\begin{table}[ht]
\centering
\footnotesize
\begin{tabular}{l@{\hspace{0.1cm}}l@{\hspace{0.1cm}}l}
\hline
\hline
\textbf{Symbol} & \textbf{Parameter name} & \textbf{Prior (range)} \\
\hline
$\mathcal{M}_c$     & Chirp mass & Uniform [15, 150] M$_\odot$ \\
$q$                 & Mass ratio & Uniform [0.125, 1.0] \\
$a_1$               & Primary spin magnitude & Uniform [0, 0.99] \\
$a_2$               & Secondary spin magnitude & Uniform [0, 0.99] \\
$\theta_1$          & Primary spin tilt angle & Sine [0, $\pi$] \\
$\theta_2$          & Secondary spin tilt angle & Sine [0, $\pi$] \\
$\phi_{12}$         & Spin azimuthal angle difference & Uniform [0, $2\pi$] \\
$\phi_{jl}$         & Precession phase & Uniform [0, $2\pi$] \\
$t_c$               & Coalescence time & Uniform [$-0.1$, 0.1] s \\
$d_L$               & Luminosity distance & Uniform [100, 3000] Mpc \\
$\theta$            & Inclination angle & Sine [0, $\pi$] \\
$\phi_c$            & Coalescence phase & Uniform [0, $2\pi$] \\
$\alpha$            & Right ascension & Uniform [0, $2\pi$] \\
$\delta$            & Declination & Cosine [$-\pi/2$, $\pi/2$] \\
$\psi$              & Polarization angle & Uniform [0, $\pi$] \\
\hline
$y$                 & Impact parameter & Uniform [0.1, 5] \\
$\log_{10}M_{Lz}$   & Redshifted lens mass ($\log_{10}$) & Uniform [1, 4] M$_\odot$ \\
\hline
\end{tabular}
\caption{Priors used for preparing the datasets for training \dingo. The bottom two rows indicate the lens parameters prior used to generate training dataset for \lensdingo{}. The same priors are used for the Bilby.}
\label{tab:priors}
\end{table}

\begin{figure}[!t]
\centering
\includegraphics[width = 0.5\textwidth]{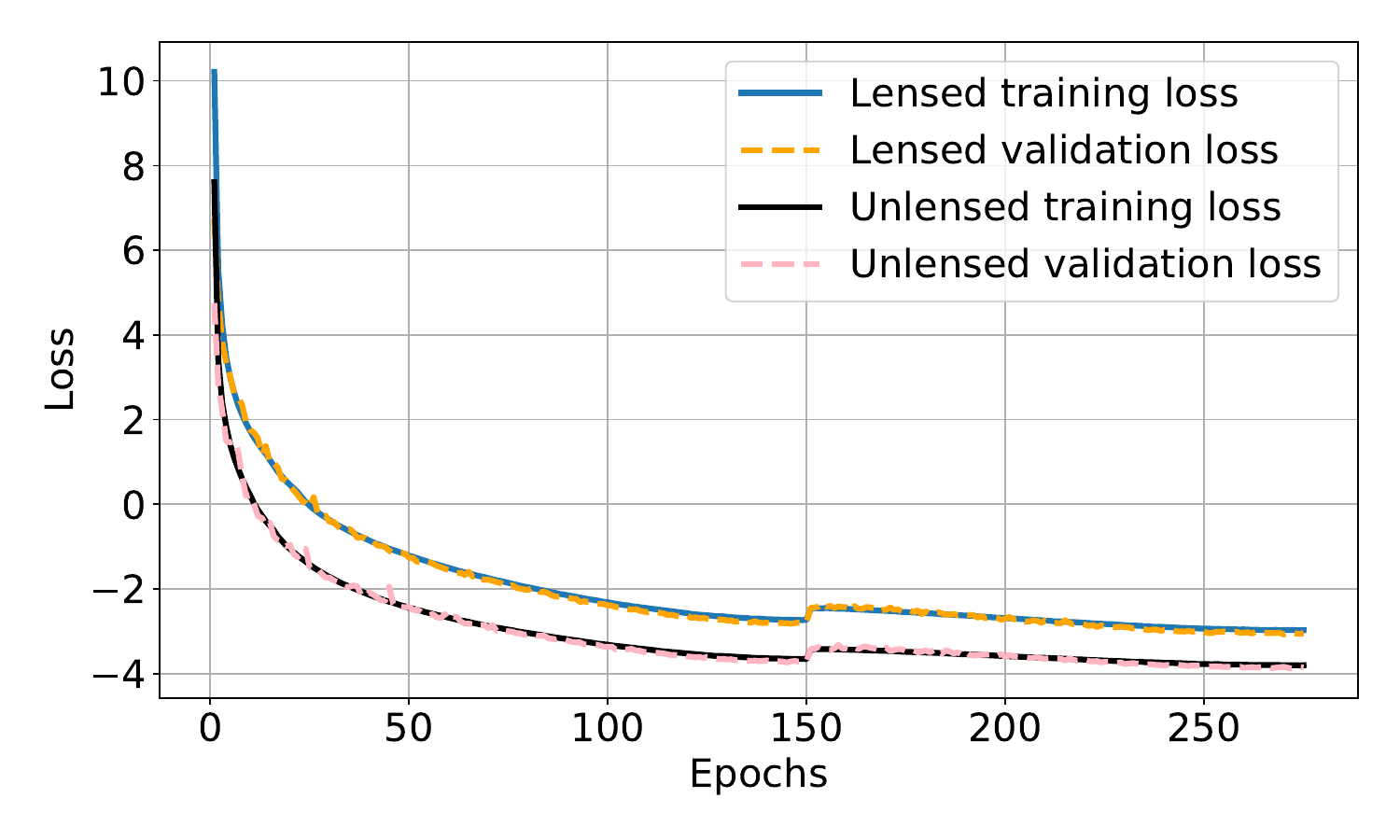}
\caption{Evolution of the training and validation loss for both
\uldingo{} and \lensdingo{} as a function of
epochs. It demonstrates similar training and validation losses, indicating no overfitting. Note that one cannot compare directly the losses of the two networks as they have different priors and number of parameters.  \label{fig:loss}}

\end{figure}

We train networks using one 40 GB NVIDIA A100 GPU, with 
16 CPUs for data preprocessing. We use a batch size of 4096 and the network architecture settings from~\cite{Dax:2021tsq}. We train both networks for 275 epochs, which takes approximately two weeks. The loss functions for both \uldingo{} and \lensdingo{} are reported in Fig.~\ref{fig:loss}.

The main computational cost lies in the training of the networks, which requires approximately two weeks to fully train the model. However, once trained, drawing $10^5$ samples from the network takes just a few minutes.

\section{Test Results}\label{sec:results}
In this section we begin by performing a statistical test called ``probability–probability'' ($p-p$) test for network validation. Then we compare the results with Bayesian PE using Bilby. Finally, we investigate why lensing inference is challenging compared to the unlensed case.

\begin{figure*}[!t]
\centering
\includegraphics[width = 0.45\linewidth]{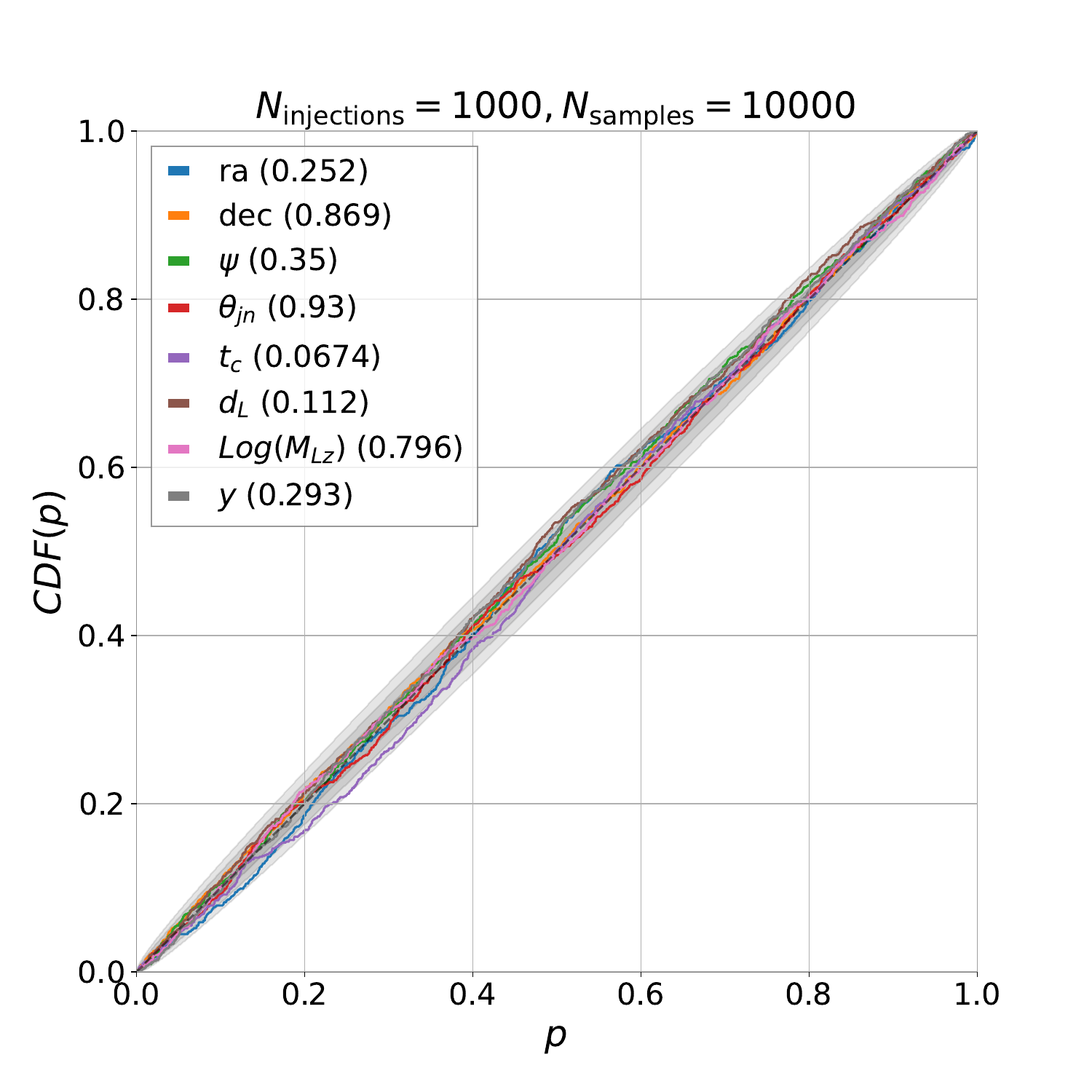}
\includegraphics[width = 0.45\linewidth]{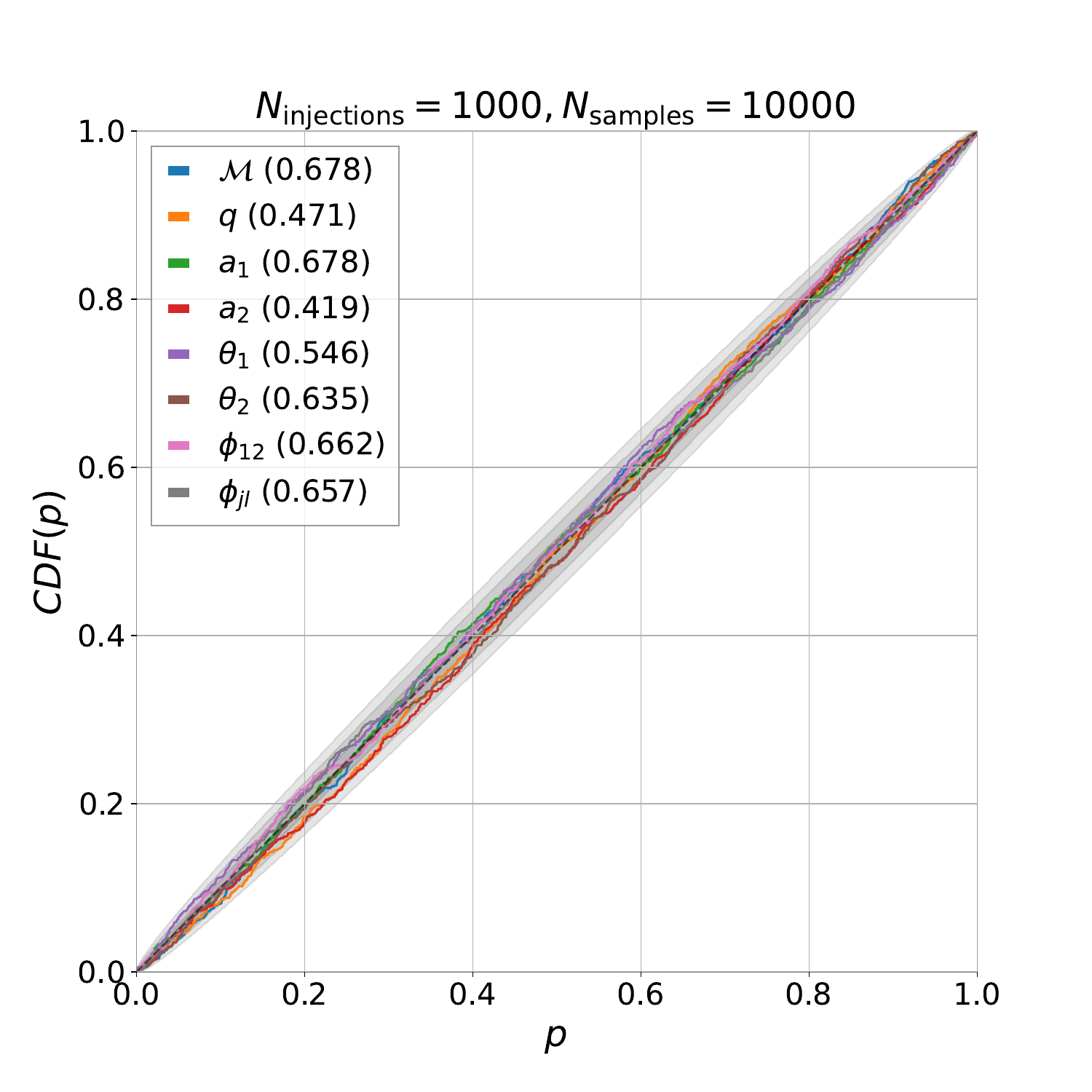}
\caption{\label{fig:pp_plots} $p-p$ plots comparing the posterior distributions predicted by DINGO without IS. The numbers in parentheses indicate the corresponding  $p$ values associated with each curve, while $N_\text{injections}$ and $N_\text{samples}$ denote the number of simulated injections and the number of posterior samples drawn per injection, respectively. The alignment with the diagonal line indicates a good calibration of the model that is the inferred parameters fall within the predicted posterior probability regions at the correct rates. As shown by the gray shaded regions, all parameters lie within 2$\sigma$ confidence interval, except the coalescence time $t_c$, which is recovered within 3$\sigma$. We divided in extrinsic and intrinsic parameters for better visualization.}
\end{figure*}

\subsection{$p-p$ test}
To validate the \lensdingo{} network, we conduct a $p-p$ test using 
1000 lensed injections drawn from the prior in Table~\ref{tab:priors}, 
each recovered using \lensdingo{} without IS. 
For each injection, we determine the 
$p$-credible region in which the true parameter lies within the corresponding posterior. If the inference is well calibrated, these cumulative probabilities, over many injections, should follow a uniform distribution between 0 and 1. 

Figure~\ref{fig:pp_plots} displays the resulting $p-p$ plots for the \lensdingo{} network, where all parameters align closely with the diagonal $p = \mathrm{CDF}(p)$, where CDF denotes the cumulative distribution function, confirming that the network remains well calibrated even without IS. 
We note that the coalescence time $t_c$ exhibits the lowest $p$ value, as the network may be having difficulty inferring the coalescence time due to a degeneracy in lensed waveforms, which are equivalent to the interference between two time-delayed waveforms for higher lens masses. 
This results in a bimodal $t_c$ posterior sometimes, which gets resolved with increased number of GNPE iterations and after IS.  Nevertheless, all other parameters are recovered within $2\sigma$ confidence interval, indicating that
\lensdingo{} can be used effectively, even without IS, for rapid
identification of lensed events and for population studies. Note that this test has already been performed for the \uldingo{} network in previous studies~\cite{Dax:2021tsq,Dax:2022pxd}.

\subsection{Comparison with Bayesian PE}

\begin{figure}[tbh]
\centering
\includegraphics[width = \linewidth]{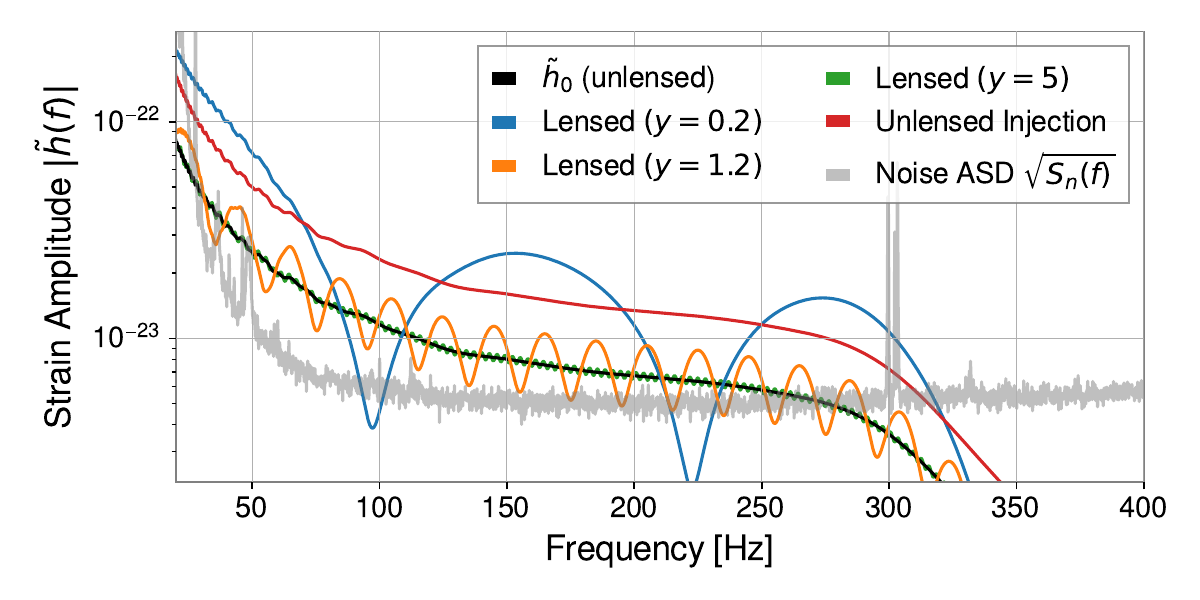}
\caption{\label{fig:waveforms_plots} Waveforms for the injected lensed and unlensed binary black hole merger are shown. The lens is modeled as a point mass with $M_{Lz}=1000 M_\odot$ and various values of $y$. Compared to the corresponding unlensed waveform (black), the lensed signals display a more complex modulation pattern because they include the WO modulation factor $F(f)$. The noise amplitude spectral density (ASD=$\sqrt{S_n(f)}$) is plotted in gray. The lensed waveform with $y=5$ (green) is very similar to the unlensed signal: although the lensing effect is present, the microlensing-induced distortions are too small to be clearly distinguished from an unlensed waveform with current detector network sensitivities. The injection parameters are listed in Table~\ref{tab:injected_params_comparison}.}

\end{figure}

\begin{table}[ht]
\centering
\footnotesize
\begin{tabular}{l@{\hspace{1cm}}l@{\hspace{1cm}}l}
\hline
\hline
Parameter & Lensed $y=0.2$ $(1.2)$ & Unlensed  \\
\hline
$\mathcal{M}_c$ & 28.10 M$_\odot$ & 28.10 M$_\odot$ \\
$q$ & 0.8 & 0.8 \\
$a_1$ & 0.4 & 0.4 \\
$a_2$ & 0.3 & 0.3 \\
$\theta_1$ & 0.5 rad & 0.5 rad \\
$\theta_2$ & 1.0 rad & 1.0 rad \\
$\phi_{12}$ & 1.7 rad & 1.7 rad \\
$\phi_{jl}$ & 0.3 rad & 0.3 rad \\
$t_c$ & 0 s & 0 s \\
$d_L$ & 1640 Mpc & 820 Mpc \\
$\theta$ & 0.4 rad & 0.4 rad \\
$\phi$ & 1.3 rad & 1.3 rad \\
$\alpha$ & 1.375 rad & 1.375 rad \\
$\delta$ & 1.121 rad & 1.121 rad \\
$\psi$ & 2.659 rad & 2.659 rad \\
$y$ & 0.2 (1.2) & -- \\
$\log(M_{Lz})$ & 3.0 & -- \\
\hline
Optimal SNR & 35.7 (18.1) & 32 \\
\hline
\end{tabular}
\caption{Parameters corresponding to lensed and unlensed injections used for testing DINGO and comparing with Bilby.}
\label{tab:injected_params_comparison}
\end{table}

We compare the results of \lensdingo{} and \uldingo{} with Bayesian PE using Bilby for a simulated black hole binary merger injected into Gaussian noise.  We particularly focus on the following three cases: (i) an \textit{unlensed injection} (GW150914-like) , (ii) \textit{lensed injection} with $y=1.2$, $M_{Lz} = 1000 M_\odot$ and  (iii) \textit{lensed injection} with $y=0.2$ and $M_{Lz} = 1000 M_\odot$. The full parameters for these injections are listed in Table~\ref{tab:injected_params_comparison}. 

Figure~\ref{fig:waveforms_plots} shows the waveform corresponding to these injections. As seen in the figure, the lensed signals are much richer than the corresponding base (unlensed) waveform $\tilde h_0(f)$ for smaller impact parameters and, for $y=5$, the lensing modulations are almost undetectable. Note that the unlensed injection waveform is the same as $\tilde h_0(f)$ but with a smaller luminosity distance, and hence higher amplitude.  

\begin{table*}[tbh]
\centering
\renewcommand{\arraystretch}{1.3} 
\setlength{\tabcolsep}{10pt} 
\begin{tabular}{l|c|c|c}
\toprule
 & \textbf{Unlensed} & \textbf{Lensed ($y = 1.2$)} & \textbf{Lensed ($y = 0.2$)} \\
\hline
\lensdingo{} & $\log Z = 497.1$  & $\log Z = 133.9$ & $\log Z = 592.3$ \\
 & $\epsilon = 0.11\%$, $n_{\rm eff} = 1110$  & $\epsilon = 0.74\%$, $n_{\rm eff} = 7442$ & $\epsilon = 0.06\%$, $n_{\rm eff} = 602$ \\
\hline
Bilby $\log Z_L$ & 498.9  & 134.7  &  593.8 \\
\hline
\hline
\hline

\uldingo{} & $\log Z = 497.9$  & $\log Z = 119.9$ & $\log Z = 467.4$ \\
 & $\epsilon = 1.8\%$, $n_{\rm eff} = 17936$  & $\epsilon = 10.3\%$, $n_{\rm eff} = 103233$ & $\epsilon = 0.0001\%$, $n_{\rm eff} = 1$ (unreliable) \\
\hline
Bilby $\log Z_U$ & 499.7 & 121.45  & 482.9 \\
\hline
\hline
\hline

DINGO $\log B^L_U$ & $-0.8$ & $14.0$ & $124.9$ \\
\hline
Bilby $\log B^L_U$ & $-0.8$ & $13.25$ & $110.9$\\
\hline
\hline
\end{tabular}
\caption{Comparison of log evidence under lensed and unlensed hypotheses between DINGO-IS and Bilby for the three injections. The last two rows compare the Bayes factors $\log B^L_U$. All cases agree well with Bilby, except for the lensed injection with $y=0.2$ analyzed with \uldingo{}. In this case, the data are out of distribution for the unlensed network, causing the sampling efficiency to drop to $\epsilon \sim 0.0001\%$ with an effective sample size of $n_{\rm eff}=1$, rendering the DINGO Bayes factor unreliable. However, for cases where the effective sample size is sufficient (e.g., $y=1.2$), DINGO Bayes factors match Bilby well and allow for fast background distribution computation. \label{tab:results}}
\end{table*}

For each of these injections, we perform  PE using \lensdingo{}, \uldingo{} and with Bilby under lensed and unlensed hypothesis ($H_L$ and $H_U$). We also perform IS for DINGO posteriors, generating $n=10^6$ samples to get improved results (DINGO-IS), while computing sampling efficiency $\epsilon$, effective sample size $n_{\rm eff}$, and log evidence  $\log Z$. We also compare log evidence and the Bayes factors $B^L_U$ between DINGO-IS and Bilby for the three injections, which is summarised in Table~\ref{tab:results}. We now discuss our findings for each of these cases in detail.\\

\begin{figure}[t!]
\includegraphics[width=\linewidth]{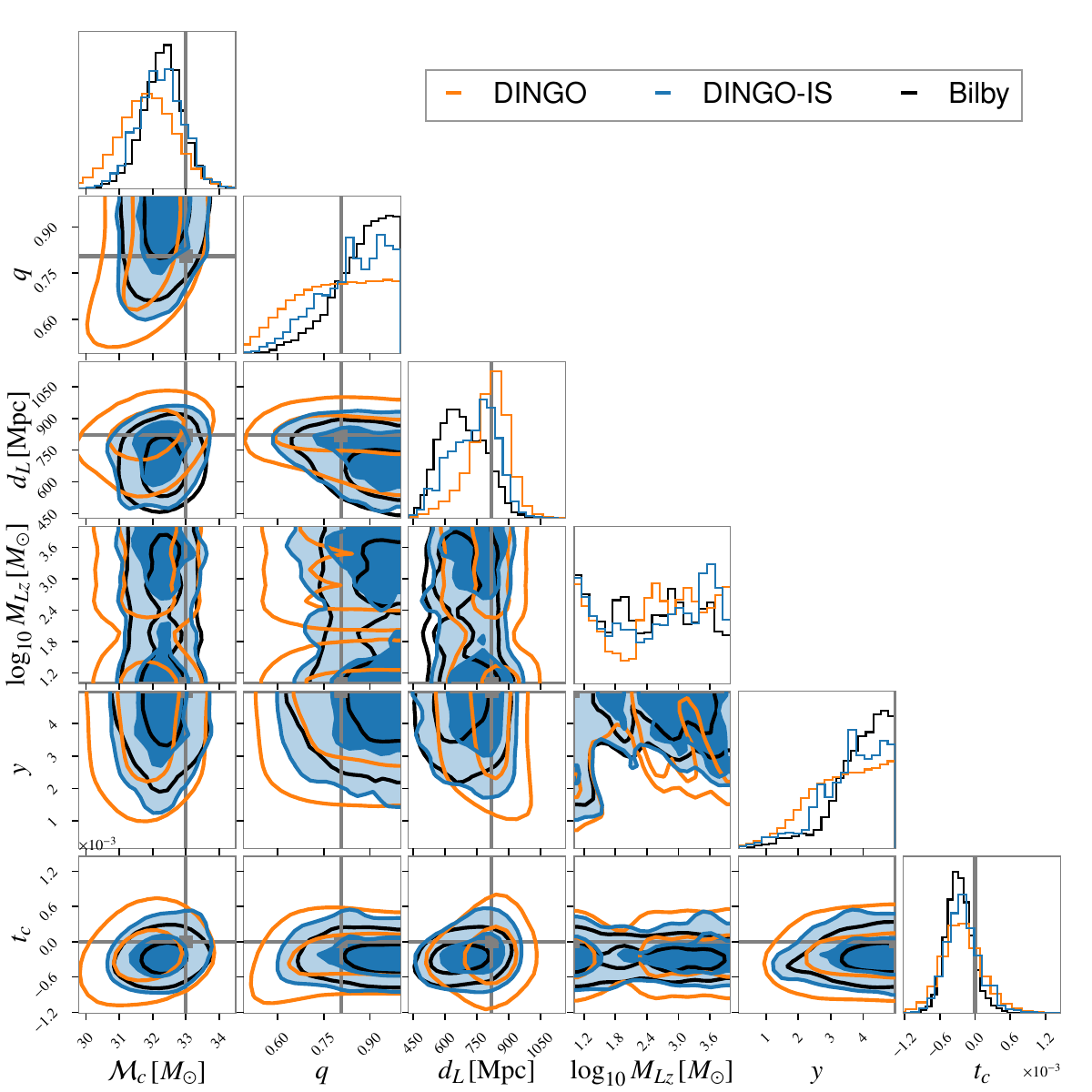}
\caption{\label{fig:ul_inj_lensed_rec} \textbf{Unlensed injection with \lensdingo{} and Bilby under lensed hypothesis}. Posterior \footnote{We do not show spin angles due to size contraints, even though we infer them.}{ distributions} from both DINGO (orange) and DINGO-IS (blue) rail at the prior boundary of the impact parameter $y=5$, and nearly flat in $\log M_{Lz}$ and recover the source parameters (gray). For this injection, we obtain $\epsilon=0.11\%$, $n_{\rm eff}=1110$, and \lensdingo{} $\log Z=497.1$ (consistent with Bilby $\log Z_L=498.9$).} 
\end{figure}

\noindent (i) \textit{Unlensed injection}: Figure~\ref{fig:ul_inj_lensed_rec} shows the posteriors obtained with \lensdingo{} and Bilby, which agree closely. As expected, $y$ saturates at its prior maximum ($y = 5$), while $\log_{10}(M_{Lz})$ remains unconstrained. This behavior occurs because, for large impact parameters, the lensed waveform becomes nearly indistinguishable from the unlensed one; in particular, for $y = 5$ the two are indistinguishable, as illustrated in Fig.~\ref{fig:waveforms_plots}. 

Thus, the marginalised posterior probability $p(y = 5 | d)$ inferred with \lensdingo{} can be used as a fast diagnostic to flag potential lensed events, by imposing a selection criterion of the form $p(y = 5 | d) < \text{threshold}$. The effectiveness of this classification strategy can be assessed using a large set of lensed and unlensed injections, which we defer to future work. The other parameters are recovered consistently with their injected values. In this case, the sampling efficiency is $\epsilon \sim 0.11\%$. 

For the same unlensed injection analysed with \uldingo{}, we obtain a sampling efficiency of $\epsilon = 1.8\%$, together with good agreement with the Bilby posteriors (see Fig.~\ref{fig:unlensed_inj_unlensed_rec} in Supplementary Material), consistent with previous studies~\cite{Green:2020dnx, Gupte:2024jfe, Dax:2021myb}. Since for unlensed injections both \lensdingo{} and \uldingo{} perform well and yield reliable estimates of $\log Z$ and $\log B^L_U$ (see Table~\ref{tab:results}), DINGO is particularly well suited for constructing the background distribution of Bayes factors needed to assess the significance of candidate lensed events, a task for which traditional methods such as Bilby are computationally very expensive.\\ 

\begin{figure*}[tbh]
\includegraphics[width=\textwidth]{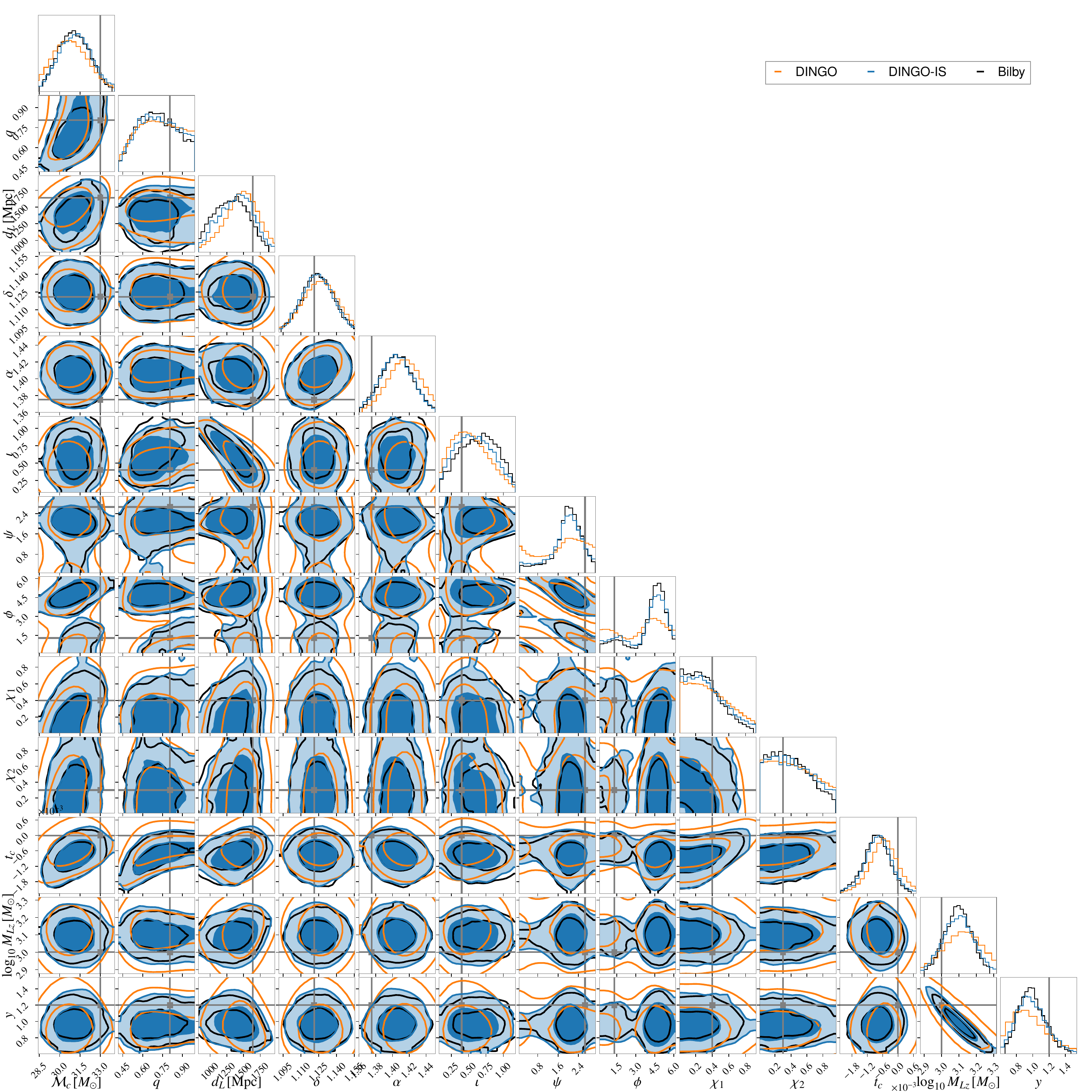}
\caption{\label{fig:lensed_inj_lensed_rec_y12} \textbf{Lensed ($y=1.2$) injection with \lensdingo{} and Bilby under lensed hypothesis}. DINGO posteriors match to that of Bilby, especially after IS while recovering the injection (gray). We obtain, $\epsilon=0.74\%$, $n_{\rm eff}=7442$, and \lensdingo{} $\log Z=133.9$ which is consistent with Bilby $\log Z_L=134.7$.}
\end{figure*}

\noindent (ii) \textit{Lensed injection} ($y=1.2$): In Fig.~\ref{fig:lensed_inj_lensed_rec_y12}, the \lensdingo{} posteriors agree closely with those obtained using Bilby, especially after applying IS, and they recover the injected parameters. The sampling efficiency in this case is $\epsilon \sim 0.74\%$.

\begin{figure}[!tp]
\centering
\includegraphics[width=\linewidth]{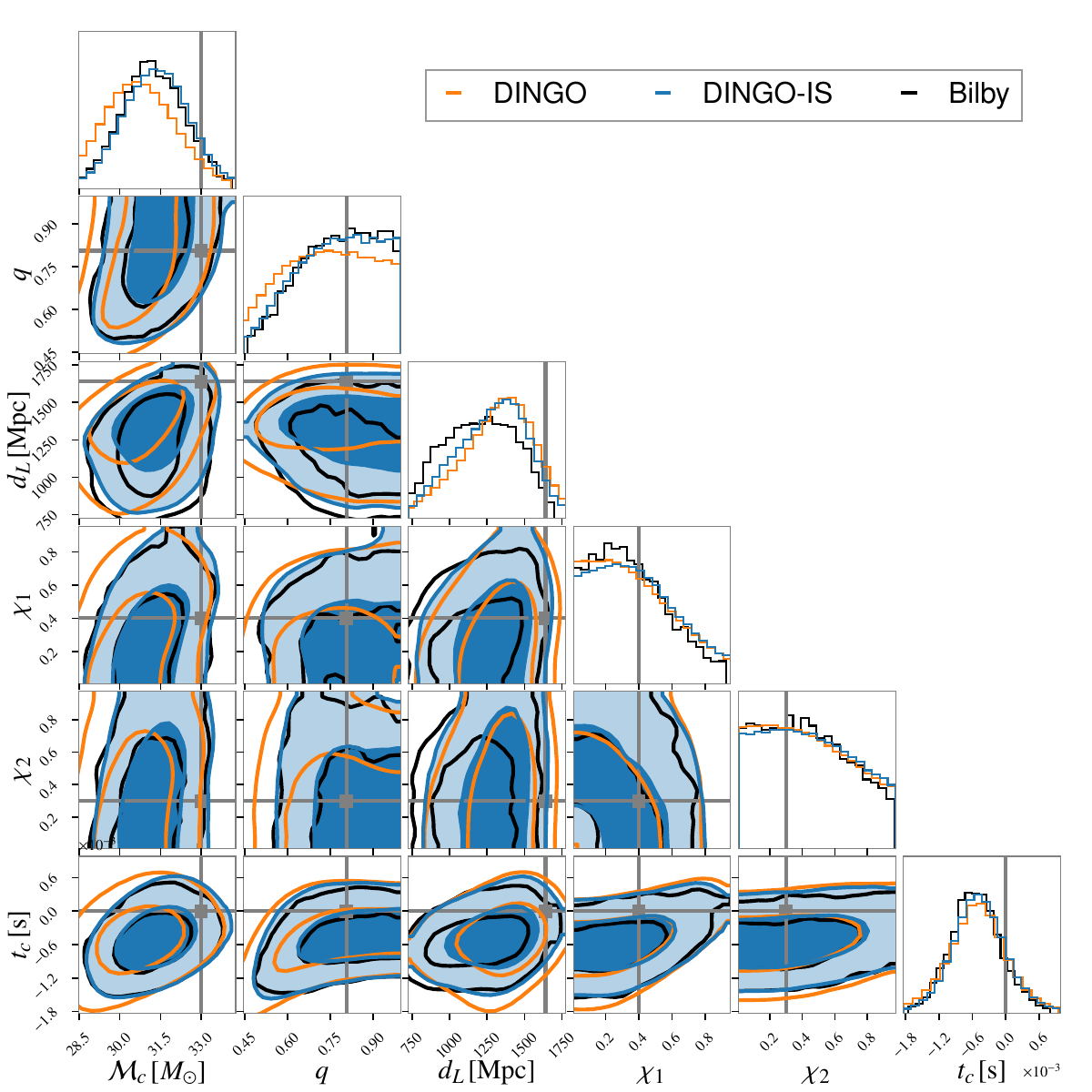}\\
(a) Lensed ($y=1.2$) injection with \uldingo{}\\
\includegraphics[width=\linewidth]{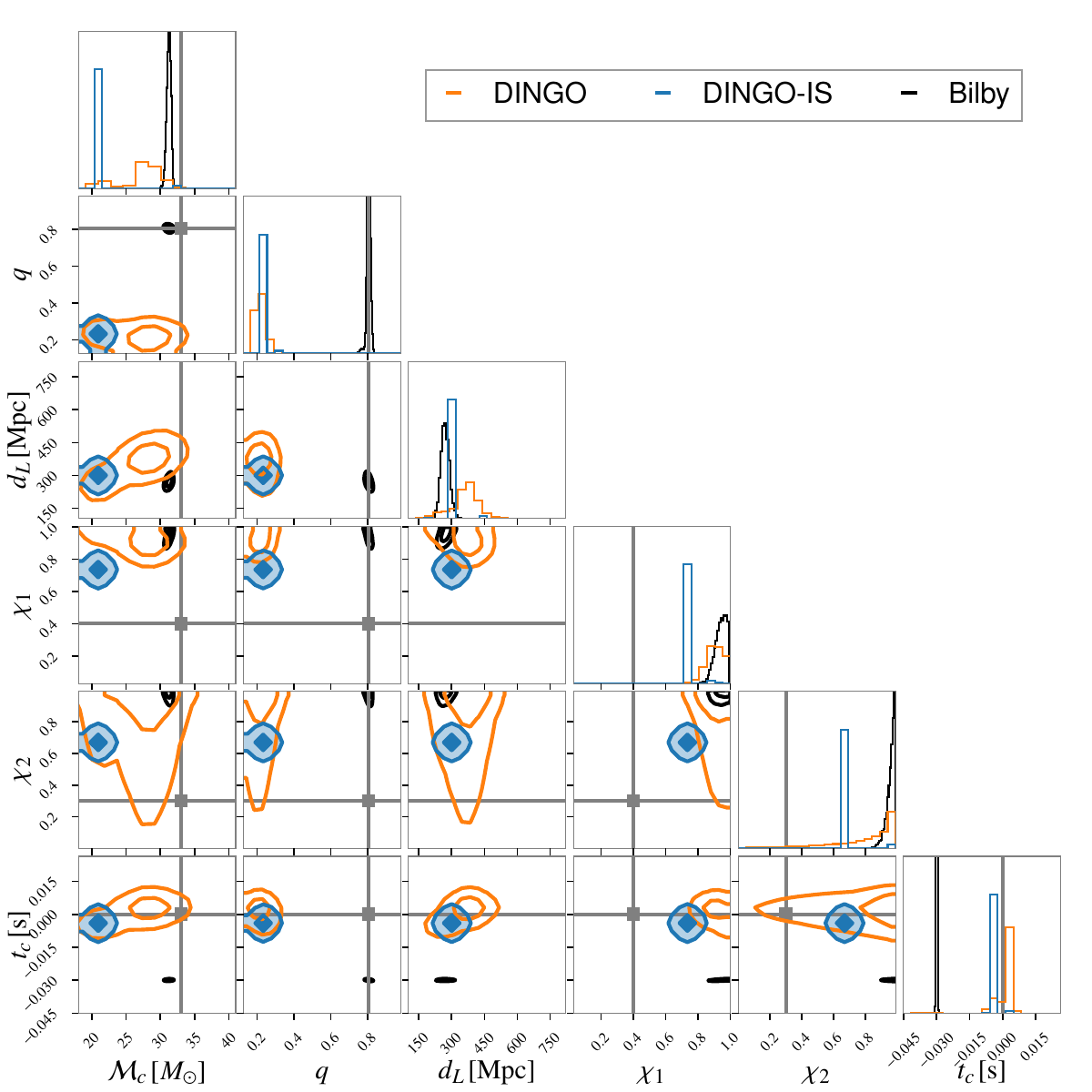}\\
(b) Lensed ($y=0.2$) injection with \uldingo{}\\
\caption{\label{fig:lensed_inj_ul_rec}  \textbf{Lensed injections with \uldingo{} and Bilby under unlensed hypothesis.} (a): For lensed ($y=1.2$) injection, the \uldingo{} network performs well with high sampling efficiency $\epsilon=10.3\%$ ($n_{\rm eff}=103233$), yielding $\log Z=119.9$ which matches Bilby $\log Z_U=121.45$. (b): For lensed ($y=0.2$) injection, the strong lensing features effectively create out-of-distribution data for the unlensed network. The sampling efficiency drops to $\epsilon=0.0001\%$ ($n_{\rm eff}=1$), resulting in unreliable posteriors and evidence estimates ($\log Z=467.4$ vs Bilby $\log Z_U=482.9$). However, this low efficiency serves as a diagnostic flag for potential lensing. Notice that here the true (Bilby) posteriors are biased more from the injection (gray) compared to the above case because of the unaccounted microlensing signatures, as expected.}
\end{figure}

The same injection with the \uldingo{} network also performs well and agrees with Bilby, achieving a sampling efficiency of $\epsilon \sim 10.3\%$. A subset of its posterior is displayed in Fig.~\ref{fig:lensed_inj_ul_rec}(a). 
Thus, for this particular lensed injection, we can obtain reliable Bayes factors, $B^L_U$. However, as we will show in the next case, this does not always hold.\\

\begin{figure*}[tbh]
\includegraphics[width=\textwidth]{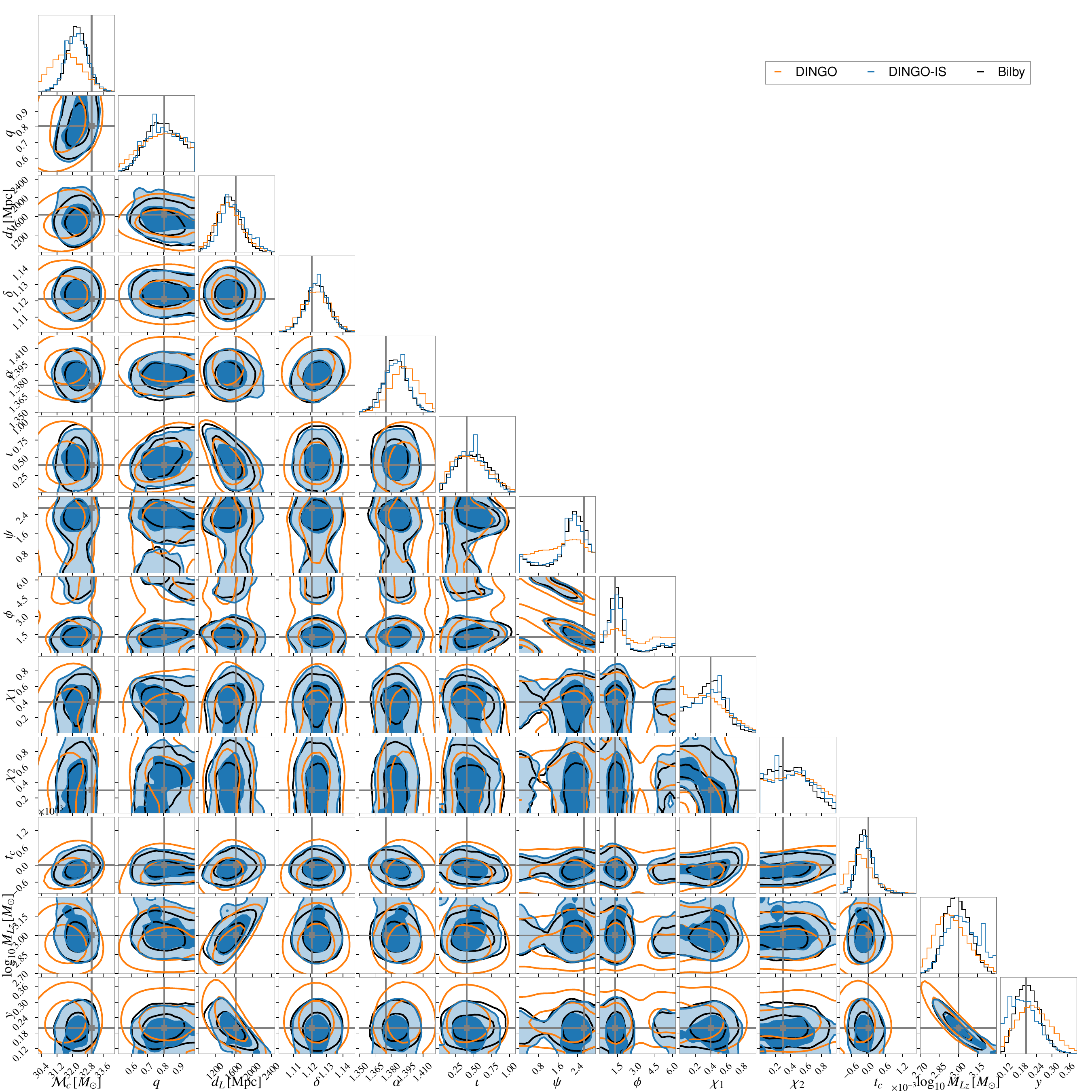}
\caption{\label{fig:lensed_inj_lensed_rec_y02} \textbf{Lensed ($y=0.2$) injection with \lensdingo{} and Bilby under lensed hypothesis}. DINGO posteriors match that of Bilby, especially after IS while recovering the injection (gray). We obtain $\epsilon=0.06\%$, $n_{\rm eff}=602$, and \lensdingo{} $\log Z=592.3$  which is consistent with Bilby $\log Z_L=593.8$.}
\end{figure*}

\noindent (iii) \textit{Lensed injection} ($y=0.2$): Here the lensing-induced modulation is stronger than in the previous case. In Fig.~\ref{fig:lensed_inj_lensed_rec_y02}, the \lensdingo{} posteriors, and therefore the corresponding values of $\log Z$, are consistent with those obtained using Bilby. For this analysis, we measure a sampling efficiency of $\epsilon \sim 0.06\%$.

In contrast, when we analyse the same injection with \uldingo{}, the sampling efficiency plummets to $\sim 0.0001\%$, and the effective sample size is reduced to $n_{\rm eff}=1$. This behaviour arises because the signal is effectively ``out of distribution'' relative to the network’s training set. A subset of the inferred posteriors, displayed 
in Fig.~\ref{fig:lensed_inj_ul_rec}(b), shows that the DINGO and DINGO-IS posteriors peak at different parameter values, and neither agrees with the Bilby posterior. This demonstrates that, for lensed injections analysed under the unlensed hypothesis, both the inferred posteriors and the evidence become unreliable (see Table~\ref{tab:results}) once the sampling efficiency is very low. 

Nonetheless, the pronounced drop in efficiency still serves as a useful diagnostic to flag potential out-of-distribution events, such as lensed signals or instrumental artifacts. We emphasise that such an efficiency drop will not always occur, particularly for weakly lensed events or those with low SNR, as in the previous example, where the unlensed hypothesis did not introduce substantial parameter biases. Consequently, for foreground (lensed) events, one must exercise caution when using DINGO to compute Bayes factors. Reliable identification of candidate lensed events requires that the sampling efficiency and effective sample size be sufficiently high (e.g., $n_{\rm eff}>500$). \\

In summary, we find that after IS, the log-evidence values and posterior distributions for all scenarios (the three injections analysed with both lensed and unlensed DINGO networks) are in good agreement with Bilby, with one exception: the lensed injection with $y=0.2$ analysed using \uldingo{}, where the pronounced reduction in sampling efficiency clearly indicates out-of-distribution data. We also notice that on an average the sampling efficiencies of \lensdingo{} are lower than the \uldingo{} (see Table \ref{tab:results}), which can be attributed to the low compressibility of the lensed waveforms. We point the reader to Appendix~\ref{app:svd} for details.

To assess the sampling time between Bilby and \lensdingo{} we can compare the number of likelihood evaluations from each sampler. Given that our sample efficiency ranges from $0.06\%$ to $0.74\%$ in Table~\ref{tab:results}, and if we require $5000$ effective samples, we estimate we need $6.8 \times 10^5$–$8.3 \times 10^6$ likelihood evaluations per event. This is doubled due to the phase reconstruction step, bringing the total to $1.4 \times 10^6$–$1.7 \times 10^7$ evaluations. For the runs in Table~\ref{tab:results} Bilby computes $6.5 \times 10^7$ likelihood evaluations. Therefore, we obtain an $\sim$$4\times$ to $\sim$$48\times$ [i.e. $\mathcal{O}(10\times)$] speedup in total CPU time. We note that while this comparison is in terms of CPU time, the likelihood evaluations for DINGO can be parallelised across nodes, further reducing the wall-clock time. We note that DINGO also has the GPU sampling time, which is $\mathcal{O}(\text{h})$ per event for generating $10^6$ proposal samples. In practice for the injections above under lensed hypothesis, the Bilby run took $\sim$$4.5$ days on four CPU cores for us, whereas DINGO-IS gave the result in less than $2$ h using hyperparallelisation. Overall, DINGO-IS is substantially faster than Bilby and will be advantageous for efficiently analysing large numbers of events in forthcoming observing runs. Note that the DINGO is expected to remain numerically stable even at SNRs lower than those studied here and can, in fact, more readily capture the broader posteriors that arise in such situations. However, in this regime microlensing signatures are anticipated to become increasingly indistinguishable from detector noise, resulting in posteriors that are largely dominated by the prior.


\section{Conclusions} \label{sec:concl} 
GW lensing provides a novel window into cosmology and astrophysics, enabling tests of dark matter, constraints on cosmological parameters, and insights into the Universe’s history. However, extracting lensing signatures can be challenging due to the lens complexity and high computational cost of WO inference. 
As a proof of concept, in this work we implemented lensing modifications in DINGO through \glow{} and trained \lensdingo{} network to infer parameters for an isolated point-mass lens. The \lensdingo{} network can serve as an efficient alternative for microlensing PE and is well calibrated as per the $p-p$ test,
see Fig.~\ref{fig:pp_plots}. It can generate $\sim 10^5$ samples/min (without IS) and enable real-time flagging of potentially microlensed events by monitoring the posterior $p(y|d)$. After IS, one can estimate the evidence and the posteriors are more reliable, consistent with Bilby. Depending on the sampling efficiency, the inference is $\mathcal{O}(10\times)$ faster than Bilby. Hence, \lensdingo{} can be utilised to study a large population of microlensed events.

Analysing microlensed signals with \uldingo{} network may sometimes result in a dramatic drop in sampling efficiency ($\epsilon \sim 10^{-6}$), leading to unreliable posteriors and evidence of Bayes-factor estimates. This serves as a useful diagnostic tool to flag out-of-distribution events, such as microlensed signals or instrumental artifacts. Since both \lensdingo{} and \uldingo{} give reliable posteriors for unlensed events, and are substantially faster than Bilby, they are uniquely suited for computing the background distributions of Bayes factors. This is useful for efficient estimation of the statistical significance of candidate microlensed events across thousands of noise realisations~\cite{junos231123}. Note that our \lensdingo{} network achieves comparatively lower sampling efficiencies than \uldingo{} and those reported in earlier works~\cite{Gupte:2024jfe, Dax:2021myb}, potentially due to the low compressibility of lensed waveforms, see Appendix~\ref{app:svd}. To enhance the performance of \lensdingo{} in the future, one could either simplify the input waveforms by calibrating the arrival time of the second micro-image using GNPE or increase the complexity of the embedding network \cite{kofler2025flexiblegravitationalwaveparameterestimation}.
    
Throughout this study, we considered simulated GW signals embedded in stationary Gaussian noise. Although no microlensing candidates were identified during the third observing run of LVK, future work will focus on extending this analysis to real events. The DINGO framework has already been successfully deployed to analyse real events, and with the methodology developed here, future work will involve reanalysis of potential microlensed candidates, such as GW231123 in the fourth observing run by retraining with updated noise conditions~\cite{Goyal:2025eqo,LIGOScientific:2025cwb}. 

Overall, this study demonstrates that combining machine learning with microlensing models provides an efficient and reliable framework for rapid inference, paving the way for real-time analysis and large-scale population studies of microlensed events. In the future, this can be expanded to include more complex lens models (e.g.~lenses embedded in external potentials~\cite{Goyal:2025eqo,Shan:2025dcd}), orbital effects (e.g., eccentricity), and degeneracies with other phenomena (e.g., overlapping signals~\cite{Hu:2025vlp,Rao:2025poe}). Additionally, the current DINGO and \lensdingo{} implementations are restricted to scenarios in which the lensed signal can be treated as a single, stationary data segment in the frequency domain. Strong lensing configurations involving resolvable or partially resolvable time delays between multiple images, such as cases where signals overlap only partially or arrive as distinct events, may require a time-domain treatment \cite{gabbard} and therefore lie beyond the scope of the present framework. 
Recently, Ref.~\cite{junos} carried out a related analysis applying DINGO to lensed events, but modelling the lensing through the GO approximation as overlapping signals. Since the GO approximation is appropriate for higher lens masses (and $w \gg 1$, see Fig.~\ref{fig:amplif_factor}), whereas we restrict the lens mass to $M_{Lz}=10^4 M_\odot$, the two studies are complementary.

As the cost of evaluating the WO lensing amplification factor rises in more complex lens systems, machine learning can facilitate in-depth analysis using multiple lens models and waveform approximants.

\begin{acknowledgments}
    We are very grateful to Jonathan Gair for valuable discussions and Aditya Sharma for careful reading of the draft.
    M.C. gratefully acknowledges Savvas Nesseris and Sachiko Kuroyanagi for the opportunity to undertake a research stay at the Max Planck Institute for Gravitational Physics (Albert Einstein Institute) in Potsdam, for valuable discussions, and for helpful comments on the manuscript.
    M.C. acknowledges support from the ``Ramón Areces" Foundation through the ``Programa de Ayudas Fundación Ramón Areces para la realización de Tesis Doctorales en Ciencias de la Vida y de la Materia 2023" and from the research project PID2021-123012NB-C43 and the Spanish Research Agency (Agencia Estatal de Investigaci\'on) through the Grant IFT Centro de Excelencia Severo Ochoa No CEX2020-001007-S, funded by MCIN/AEI/10.13039/501100011033. M.C. also thanks the Max Planck Institute for Gravitational Physics for warm hospitality during the training of this project and the Programa CSIC ``iMOVE 2024'' which made this stay possible. S.R.G. is supported by a UKRI Future Leaders Fellowship (Grant No. MR/Y018060/1). This material is based upon work supported by NSF's LIGO Laboratory which is a major facility fully funded by the National Science Foundation.
    The computational work for this manuscript was carried out on the computing cluster Saraswati at the Max Planck Institute for Gravitational Physics in Potsdam.
\end{acknowledgments}

\appendix
\section{Compressibility of lensed waveforms}
\label{app:svd}
\begin{figure}[!t]
\centering
\includegraphics[width = 0.48\textwidth]{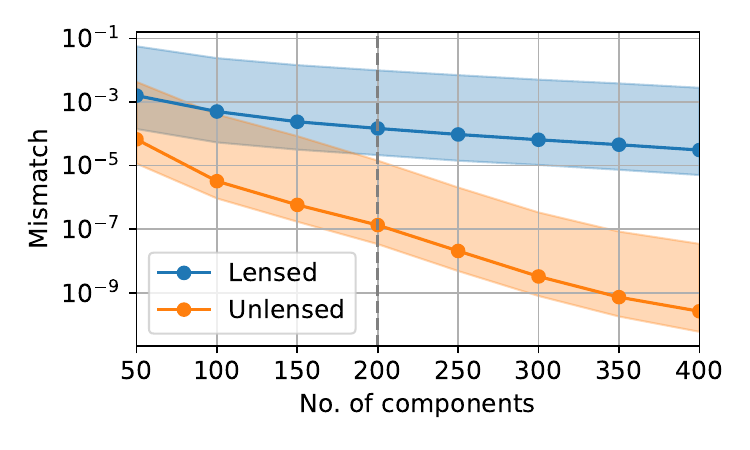}
\caption{\label{fig:svd} Mismatch between true waveforms and their reconstruction using a truncated SVD basis. Microlensed waveforms require significantly more components than unlensed ones to reach the same mismatch, reflecting their increased intrinsic complexity. The SVD is used here only as a diagnostic and not in the DINGO analysis.}
\end{figure}

Lensed waveforms are intrinsically more complex than unlensed ones. To quantify the relative complexity of unlensed and microlensed waveforms, we perform a singular value decomposition (SVD) of ensembles of time-aligned frequency-domain signals. Given a data matrix $\mathbf{H}$ whose rows correspond to individual waveforms, the decomposition \cite{Hansen1987},
\begin{equation}
\mathbf{H} = \mathbf{U}\,\mathbf{\Sigma}\,\mathbf{V}^\dagger
\end{equation}
defines an orthonormal basis $\{v_k(f)\}$ in waveform space (columns of $\mathbf{V}$). Here $\mathbf{U}$ contains the left singular vectors describing how individual waveforms project onto this basis, while $\mathbf{\Sigma}=\mathrm{diag}(\sigma_1,\sigma_2,\ldots)$ contains the singular values ordered by decreasing importance. Truncating the expansion to the first $n$ components yields an approximate reconstruction
\begin{equation}
\tilde h^{(n)}(f) = \sum_{k=1}^{n} c_k\, v_k(f),
\end{equation}
where $c_k$ denote the coefficients. The accuracy of this representation is quantified using the mismatch between $h(f)$ and $h^{(n)}(f)$. 
As shown in Fig.~\ref{fig:svd}, unlensed waveforms reach mismatch $\mathcal{M}\sim10^{-4}$ with $n\simeq200$ basis vectors, whereas microlensed waveforms require approximately twice as many components. Although DINGO does not employ an explicit SVD representation, this diagnostic highlights the increased intrinsic complexity and reduced compressibility of microlensed signals.


One can gain intuition about the reason for the worse scaling of basis vectors for the lensed case with the following argument. 
We know that not setting the merger time to the time of arrival in the detectors (i.e. the merger occurs at $t \neq 0$) drastically increases the number of SVD bases needed to  model the waveform, which is one of the reasons for using GNPE~\cite{Dax:2021myb}. If one thinks of lensing as a superposition of two signals, even though we have time aligned the first signal, the second is free to shift depending on the value of $M_{Lz}$ and $y$. While the first signal is standardised, the second is not and the network/SVD has to represent this time shift on its own. For future work, one could consider applying further time shifts to simplify the signal depending values of $M_{Lz}$ and $y$ or increase the complexity of the embedding network that DINGO uses to compress a waveform before passing it to the normalizing flow.

\clearpage
\twocolumngrid

\bibliographystyle{apsrev4-1}
\bibliography{gw_lensing}

\clearpage
\onecolumngrid
\begin{center}
    \textbf{\Large Supplementary Material}
\end{center}

\begin{figure}[h]
\centering
\includegraphics[width=\textwidth]{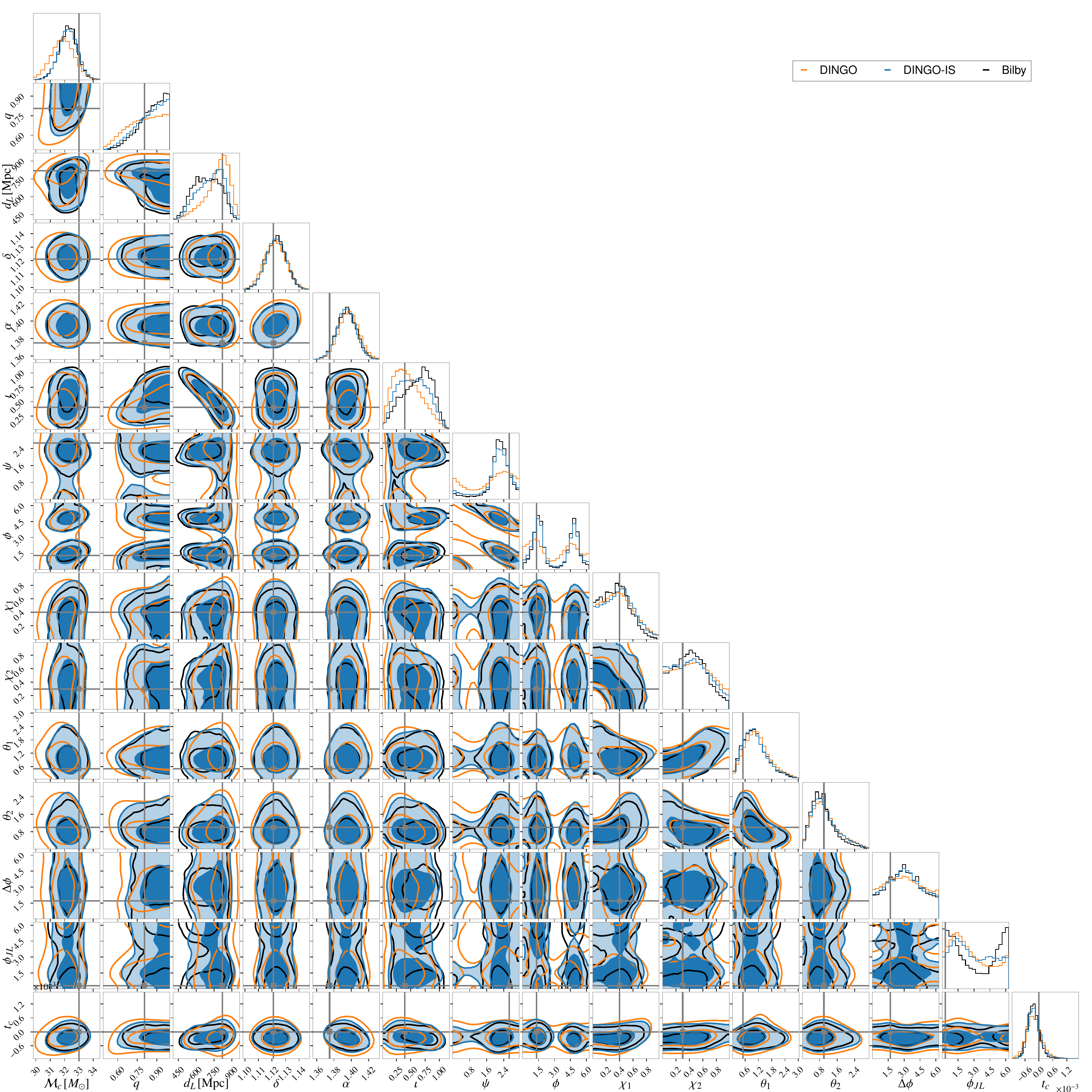}
\caption{\label{fig:unlensed_inj_unlensed_rec} \textbf{Unlensed injection analyzed with \uldingo{} and Bilby under unlensed hypothesis}. The posteriors match well between DINGO and Bilby and injection (gray) is recovered. With \uldingo{}, we obtain, $\epsilon=1.8\%$, $n_{\rm eff}=17936$,  $\log Z=497.9$ which is in agreement to Bilby $\log Z_U=499.7$.}
\end{figure}

\end{document}